\documentclass[letterpaper,twocolumn,superscriptaddress,
	    nobibnotes,aps,prb,longbibliography,floatfix]{revtex4-1}

\usepackage{amsmath, amssymb, amsfonts}
\usepackage{amstext, mathrsfs, textcomp}
\usepackage{graphicx}%
\usepackage{color}

\usepackage{hyperref}

\usepackage{changes}

\usepackage[T1]{fontenc}

\def\beq{\begin{eqnarray}}
\def\eeq{\end{eqnarray}}
\def\bec{\begin{center}}
\def\enc{\end{center}}
\def\bit{\begin{itemize}}
\def\eit{\end{itemize}}

\begin{document}

\title{Local field enhancement and thermoplasmonics in multimodal Aluminum structures}

\author{Peter R. Wiecha}
\affiliation{CEMES, University of Toulouse and CNRS (UPR 8011), 29 rue Jeanne Marvig, BP 94347, 31055 Toulouse, France}
\author{Marie-Maxime Mennemanteuil}
\affiliation{Laboratoire Interdisciplinaire Carnot de Bourgogne, CNRS UMR 6030, Universit\'e Bourgogne Franche-Comt\'e, 9 Av. A. Savary, BP 47870, 21078 Dijon, France}
\author{Dmitry Khlopin}
\affiliation{Laboratoire de Nanotechnologie et d Instrumentation Optique, Institut Charles Delaunay, UMR CNRS 6281, Universit\'e de Technologie de Troyes, France}
\author{J\'er\^ome Martin}
\affiliation{Laboratoire de Nanotechnologie et d Instrumentation Optique, Institut Charles Delaunay, UMR CNRS 6281, Universit\'e de Technologie de Troyes, France}
\author{Arnaud Arbouet}
\affiliation{CEMES, University of Toulouse and CNRS (UPR 8011), 29 rue Jeanne Marvig, BP 94347, 31055 Toulouse, France}
\author{Davy G\'erard}
\affiliation{Laboratoire de Nanotechnologie et d Instrumentation Optique, Institut Charles Delaunay, UMR CNRS 6281, Universit\'e de Technologie de Troyes, France}
\author{Alexandre Bouhelier}
\affiliation{Laboratoire Interdisciplinaire Carnot de Bourgogne, CNRS UMR 6030, Universit\'e Bourgogne Franche-Comt\'e, 9 Av. A. Savary, BP 47870, 21078 Dijon, France}
\author{J\'er\^ome Plain}
\affiliation{Laboratoire de Nanotechnologie et d Instrumentation Optique, Institut Charles Delaunay, UMR CNRS 6281, Universit\'e de Technologie de Troyes, France}
\author{Aur\'elien Cuche}
\altaffiliation{Corresponding author: aurelien.cuche@cemes.fr}
\affiliation{CEMES, University of Toulouse and CNRS (UPR 8011), 29 rue Jeanne Marvig, BP 94347, 31055 Toulouse, France}



\begin{abstract}
Aluminum nanostructures have recently been at the focus of numerous studies due to their properties including oxidation stability and surface plasmon resonances covering the ultraviolet and visible spectral windows. In this article, we reveal a new facet of this metal relevant for both plasmonics purpose and photo-thermal conversion. The field distribution of high order plasmonic resonances existing in two-dimensional Al structures is studied by nonlinear photoluminescence (nPL) microscopy in a spectral region where electronic interband transitions occur. The polarization sensitivity of the field intensity maps shows that the electric field concentration can be addressed and controlled on-demand. We use a numerical tool based on the Green dyadic method to analyze our results and to simulate the absorbed energy that is locally converted into heat. The polarization-dependent temperature increase of the Al structures is experimentally quantitatively measured, and is in an excellent agreement with theoretical predictions. Our work highlights Al as a promising candidate for designing thermal nanosources integrated in coplanar geometries for thermally assisted nanomanipulation or biophysical applications.
\end{abstract} 



\maketitle

\indent 


Metallic nanostructures sustain localized and delocalized Surface Plasmon (SP) resonances when they are excited under specific conditions. These resonances are giving rise to large enhancement of the electromagnetic field, subwalength confinement, and plasmon propagation in structures with low dimensionalities \cite{Barnes2003,Atwater2005}. Owing to their remarkable optical properties, SP resonances have led to a large number of direct applications including high-precision biological sensing \cite{Homola2008}, light manipulation by metasurfaces \cite{Capasso2011}, design of integrated devices for information processing \cite{Bozhevolnyi2010}, and highly localized heat sources \cite{Govorov2006,Girard2009}.

Due to the large negative real part of the dielectric function in the visible spectrum, gold or silver, either as lithographed patterns or as colloidal nanoparticles, represent the bulk of plasmonic devices so far. Although both metals exhibit complementary features that have fostered the fast development of plasmonics as a technology, these two noble metals also suffer from drawbacks; namely bulk oxidation for silver and an important interband absorption for gold at energies above 2.25 eV. Moreover, both are expensive materials limiting their systematic and massive use in commercial systems. These considerations triggered a rising interest for non-conventional plasmonic materials \cite{Shalaev2013,Baffou2016}. In this context, while discarded for a long time for plasmonic applications due to its high losses in the red part of the visible spectrum, aluminum has recently demonstrated its potential for applications in the blue-ultraviolet energy range \cite{Halas2012,Bisio2013,Martin2013,Halas2014,Leipner2014,Halas2015,Gerard2015,Novotny2015,Gallinet2015,Tsai2015,Huang2015}. The main asset of Al over other noble metals stems mainly from a plasma frequency $\omega _{\rm p}$ situated at a higher energy. In addition to its intrinsic electronic properties, the optical response of aluminum is stable over time thanks to the formation of a self-limiting oxide layer preserving the integrity of the structure. Interestingly, the quest for new plasmonic materials has also been driven by the recent interest in photo-thermal energy conversion by metallic systems. Indeed, the strong local field enhancement associated with the plasmonic resonances goes along with an enhanced absorption in the metal due to Ohmic losses. Metal nanostructures can therefore increase the temperature in their environment and be used as integrated heat nanosources \cite{Govorov2006,Govorov2007,Richardson2009,Girard2009,Girard2010}.

\begin{figure}[htb]
\centering{
\includegraphics[width=8cm]{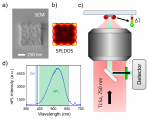}}
 \caption{(a) SEM image of an aluminum square. Scale bar represents 250 nm. (b) SP density of states computed for the same metallic pad at $\lambda$=750 nm. (c) Sketch of the nonlinear photoluminescence microscope. (d) Nonlinear emission spectrum acquired from the aluminum structure shown in (a) under pulsed excitation at $\lambda $=750 nm.}
 \label{fig1}
\end{figure}

Thanks to its ability to support surface plasmon resonances tunable in the visible to IR spectral ranges and its biocompatibility, gold has been given a central role in the studies of the conversion of surface plasmons into heat. This active research field, called thermoplasmonics, could potentially lead to breakthroughs in several fields like hyperthermia-induced apoptosis in oncology \cite{Halas01}, nanochemistry \cite{Quidant2014} or thermo-hydrodynamically assisted plasmonic trapping and manipulation \cite{Cuche2013,Toussaint2014}, where the onset of temperature gradients is critical. The power dissipated in a nanostructure is a complex interplay between the optical near-field intensity and the material properties \cite{Baffou2016,Viarbitskaya2015}. Localized SP resonances with large quality factors require a low dissipation: a compromise is therefore to be found to maximize the power absorbed inside sub-micron nanostructures. Whereas the dipolar surface plasmon resonance of subwavelength gold nanospheres partly overlaps with interband transitions, this is generally not the case for higher-order modes as retardation and shape effects shift surface plasmon resonances to the red. The contribution of interband transitions to energy dissipation is therefore suppressed in gold at infrared wavelengths where better resonance quality factors and larger field enhancements are obtained \cite{Sonnichsen2002,Bosman2013}. In aluminum on the other hand, the interband transitions are situated around $\sim$1.5 eV. The increased dissipation in this energy range can therefore be seen as a different configuration, positioning aluminum as a good candidate for temperature control experiments in the near field in a biocompatible spectral window where the cytotoxicity of Al and hyperthermia-induced apoptosis could be exploited. 

In this article, we investigate the plasmonic response and subsequent photothermal conversion of two-dimensional (2D) plasmonic aluminum cavities in the spectral window corresponding to its interband transitions ($\lambda =$750 nm). We show that these multimodal cavities sustain high order SP resonances with a two dimensional distribution which leads to a strong localization of the electric field. We performed nonlinear photoluminescence microscopy to reveal this spatial distribution for several sizes and cavity geometries. A numerical tool based on the Green Dyadic method that closely accounts for the experimental results has been used to compute the heat deposited in the Al structures. By comparing the efficiency of gold and aluminum, we demonstrate that the latter is an excellent material for thermoplasmonic applications. From this theoretical analysis, we quantify the temperature increase in the vicinity of the metallic structures as a function of the incident polarization. We finally confront these predictions to experimental data inferred by monitoring the temperature sensitivity of the electronic transport of gold nanowire placed in the vicinity of the heater.   We demonstrate a qualitative and quantitative agreement confirming the potential of Al structures for alternative and efficient integrated thermoplasmonic devices.

\section*{Results and discussion}

Unlike dipolar resonances excited in subwavelength metallic antennas, high order plasmonic resonances exhibit a large variety of spatial distributions that can be used for addressing specific regions of space below the diffraction limit. It has been shown recently that 2D plasmonic cavities made of gold or aluminum sustain these SP resonances which concentrate the electric field in subwavelength areas \cite{VanAken2011,Arbouet2013,Martin2014,Cuche2015}. We use Electron Beam Lithography (EBL) to define aluminum 2D nanostructures. Briefly, a silica coverslip is spin-coated with a layer of polymethyl methacrylate (PMMA) topped with a conductive polymer.  After exposition of the designed layout by an electron beam and subsequent development of PMMA, a 50 nm thick aluminum film is thermally deposited on the sample. The final Al structures are then obtained by removing the polymer.  Figure \ref{fig1}(a) shows a scanning electron microscope (SEM) image of a 800 nm long Al square. 

Because the nonlinear photoluminescence (nPL) response and the heat generation both rely on the amplitude of the local electric field in the metal, these types of plasmonic structures have been systematically probed by nonlinear photoluminescence microscopy \cite{VanHulst2011,Eisler2013}. We use here the generic nPL labeling since the mechanism of the nonlinear emission in metals is still being debated \cite{Lupton2015}. Such study is beyond the scope of this paper. As previously demonstrated, the all-optical nPL microscopy allows for a polarization dependent mapping of the field enhancement occurring in plasmonic structures \cite{Teulle2012}. The emitted photoluminescence can be described by the phenomenological relation:

\begin{equation}
I_{nPL}(\textbf{R}_{0},\omega) = \eta ^{2}(\omega) \int_{V} \left| \textbf{E}(\textbf{R}_{0}, \textbf{r},\omega) \right| ^{4} d\textbf{r}
\label{eq1}
\end{equation}

where $\textbf{R}_{0}$ and $\omega$ are the position and the frequency of the excitation. The prefactor $\eta$ is an effective frequency-dependent nonlinear coefficient and $V$ is the volume of the metallic particle. Finally, $\textbf{E}$ is the total electric field in the metal at the position $\textbf{r}$. Recent works have also demonstrated the intrinsic link between the nPL signal recorded from colloidal gold platelets and the underlying local plasmonic density of states in the metal \cite{Viarbitskaya2015,Cuche2015}. The band structure of Aluminum shows that electrons below the Fermi level can be excited by photons with energy close to 1.5 eV ($\lambda$ $\approx$ 800 nm), hence undergoing an interband transition at the $K$ and $W$ symmetry points \cite{Gerard2015}. On the basis of previous studies of the nPL response of Al antennas at such energies \cite{VanHulst2011}, we intentionally probed the plasmonic resonances close to the maximum of these interband transitions, as illustrated by the SP density of states (SP-LDOS) distribution computed in Fig.\ref{fig1}(b). In spite of high dissipation leading to damping of the SP resonances, the SP local density of states features maxima located along the edges of the cavity.

Figure \ref{fig1}(c) shows a sketch of the nonlinear photoluminescence microscope. nPL maps of the Al structures are acquired pixel by pixel by scanning the sample through a focused and polarization controlled laser beam from a Ti:Sapphire femtosecond laser at $\lambda=750$ nm with a pulse width of around 150 fs and repetition rate of 80 MHz. The average power at the back aperture of the objective has been kept at 1 mW (it never exceeded 2 mW). The laser-filtered photoluminescence is collected in retro-diffusion using a dichroic mirror and sent either on a sensitive photodetector or on a spectrometer with a high-sensitivity CCD. The broadband photoluminescence spectrum emitted by a 800 nm Al square is presented by the shaded area in the spectrum of Fig.\ref{fig1}(d). A weak signal at the second-harmonic frequency is also observed, but will not be discussed here.

\begin{figure}[htb]
\centering{
\includegraphics[width=8cm]{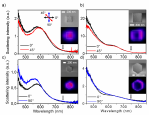}}
 \caption{(a-b) Dark-field spectra acquired on an aluminum square structure with (a) 400\,nm and (b) 2\,\textmu m side length for an excitation linearly polarized at 0 and 45$^{\circ}$. (c-d) Same as (a) and (b) for hexagonal pads for polarization at 0 and 90$^{\circ}$, respectively. SEM and DF images (with scale bars) are systematically shown in insets. The experimental excitation wavelength at 750 nm used for the nonlinear photoluminescence measurements is indicated by the gray arrows.}
 \label{fig2}
\end{figure}

We start our investigations by measuring the linear optical properties of the Aluminum structures by dark-field (DF) scattering spectroscopy. In Figure \ref{fig2}, the polarization-dependent DF-spectra acquired for two geometries and different sizes are shown. The corresponding DF images, spectrally integrated, are shown in the insets together with the respective SEM images. The scattering signal is strong around the edges and the vertices of the structures where the plasmonic resonances are localized (Fig.\ref{fig1}(b)). The full set of spectra acquired for structure sizes ranging from 300\,nm to 2\,\textmu m (Fig.S2 in reference \citenum{SI}) exhibits the same characteristics as the ones displayed in Fig.\ref{fig2}. A strong scattering response from the structures is observed in the blue with the presence of several resonances. However, for wavelengths larger than 700\,nm, a very weak scattering is observed with a systematic minimum around 800\,nm where the maximum of $Im(\epsilon)$ is observed (see Fig. S1 in reference \citenum{SI}).

\begin{figure*}[htb]
\centering{
\includegraphics[width=12cm]{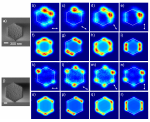}}
 \caption{(a) SEM image of a hexagonal structure with 600 nm long sides. (b-e) nPL maps acquired with incident polarization oriented at (b) 0$^{\circ}$, (c) 30$^{\circ}$, (d) 60$^{\circ}$ and (e) 90$^{\circ}$. Polarization is indicated by the white arrows.(f-i) Corresponding simulated nPL maps. (j) SEM image of a hexagonal structure with 900 nm long sides. (k-n) nPL maps acquired with polarizations similar to (b-e). (o-r) Corresponding simulated nPL maps. The scale bars are 300 nm.}
 \label{fig3}
\end{figure*}

Further insight on the SP spatial distribution can be obtained by comparing the information provided by the spectrally averaged scattering response of an Al nanostructure and the spatially resolved nPL emission at a given excitation wavelength ($\lambda$ = 750 nm, indicated by the gray arrows in the spectra of Fig.\ref{fig2}). Pixel-by-pixel nPL images collected on hexagonal structures of different sizes are shown in Fig.\ref{fig3}. As previously reported for gold crystalline platelets, such 2D cavities display a similar polarization dependent nPL signal located at specific vertices along the cavity perimeter \cite{Arbouet2013,Cuche2015,Viarbitskaya2015}. Focusing on the nPL of these objects, we observe the onset of three intense areas on the upper part of the hexagon images for a horizontal polarization (0$^{\circ}$ - along the $x$ axis). As the polarization is rotated from 0$^{\circ}$ to 90$^{\circ}$ (white arrows in Fig.\ref{fig3}), the nPL pattern also undergoes a 90$^{\circ}$ rotation with a sequential lightning of the apexes. The final configuration leads to two high intensity areas on the left and on the right part of both hexagons.

Using the expression in Eq.\ref{eq1}, we compare these results to simulated maps as shown in Fig.\ref{fig3}(f-i) and (o-r). These computations are performed with a numerical tool based on the Green dyadic formalism where a virtual focused excitation beam is raster scanned on the structure. It allows the computation of the local electric field intensity in any point of the metallic cavity, taking into account its real geometry. A detailed description of the method can be found in reference \citenum{Teulle2012}. The symmetric positions of the nPL spatial distribution and its polarization dependence are globally reproduced. Interestingly, the mirror axis defined by the paired high intensity areas experiences a clockwise evolution at this wavelength as the polarization is rotated, unlike results obtained on gold prismatic structures \cite{Viarbitskaya2013}. The similar evolution fo the nPL maps for the two cavity sizes does not straightforwardly imply that the SP resonances probed in the cavity are similar. Indeed the plasmonic landscape emerges from the interference of degenerated SP electron density waves in the cavity. Here, the difference in perimeter between both cavities is about 1.8\,\textmu m, implying that different orders of SP resonances can give rise to identical nPL signatures \cite{Arbouet2013}.

Simple Al nanostructures (0D and 1D) have been studied mainly for their plasmonic properties in the blue-UV region of the electromagnetic spectrum \cite{Halas2014,Leipner2014,Halas2015,Gerard2015}. However our observations confirm the existence of high order SP resonances in the interband spectral range in complex 2D Aluminum structures (Fig.\ref{fig3}). While these modes have a well-defined spatial distribution in the structure, the intrinsic interband transitions of the material are spatially invariant in these polycristalline Al structures. Former studies discussed the mutual interaction leading to an hybridization between the interband transitions and a plasmonic dipolar resonance promoted by the spectral overlap \cite{Lecarme2014}. However, our nPL analysis does not provide any clear information on such hybridization between the high order SP resonances and the dipole formed by an interband absorption event. 

From previous nPL investigations on gold systems, it has been established that SP resonances existing along the edges of a cavity provide an additional channel into the electronic density of states for hot electron generation \cite{Viarbitskaya2015}. The energy deposited in the electronic subsystem eventually relaxes by electron-phonon coupling, leading to a SP induced local heat generation and eventually to a partial melting of the material itself \cite{Viarbitskaya2015}. The formal link between the local electric field in the metal and the heat generated by the Joule effect at the position $\textbf{R}_{0}$ of the light beam can be defined in the following way by computing the total power $Q$ dissipated in the metal \cite{Teulle2012}:
\begin{equation}
Q (\textbf{R}_{0},\omega) = \frac{\omega}{8 \pi} Im(\epsilon) \int_{V} \left| \textbf{E}(\textbf{R}_{0}, \textbf{r},\omega) \right| ^{2} d\textbf{r}
\label{eq2}
\end{equation}
where $Im(\epsilon)$ is the imaginary part of the dielectric function of aluminum.

Eq.~\ref{eq2} shows that the photothermal conversion efficiency depends both on intrinsic losses in the metal $Im(\epsilon)$ (including interband transitions), and the local electric field intensity. While the former has an homogeneous distribution over the nanostructure, the second term has a strong spatial dependence arising from the plasmonic landscape. These two factors define the efficiency of the photothermal conversion and allow for tailoring the heat generation in such planar metallic cavities. 

\begin{figure}[htb]
\centering{
\includegraphics[width=8cm]{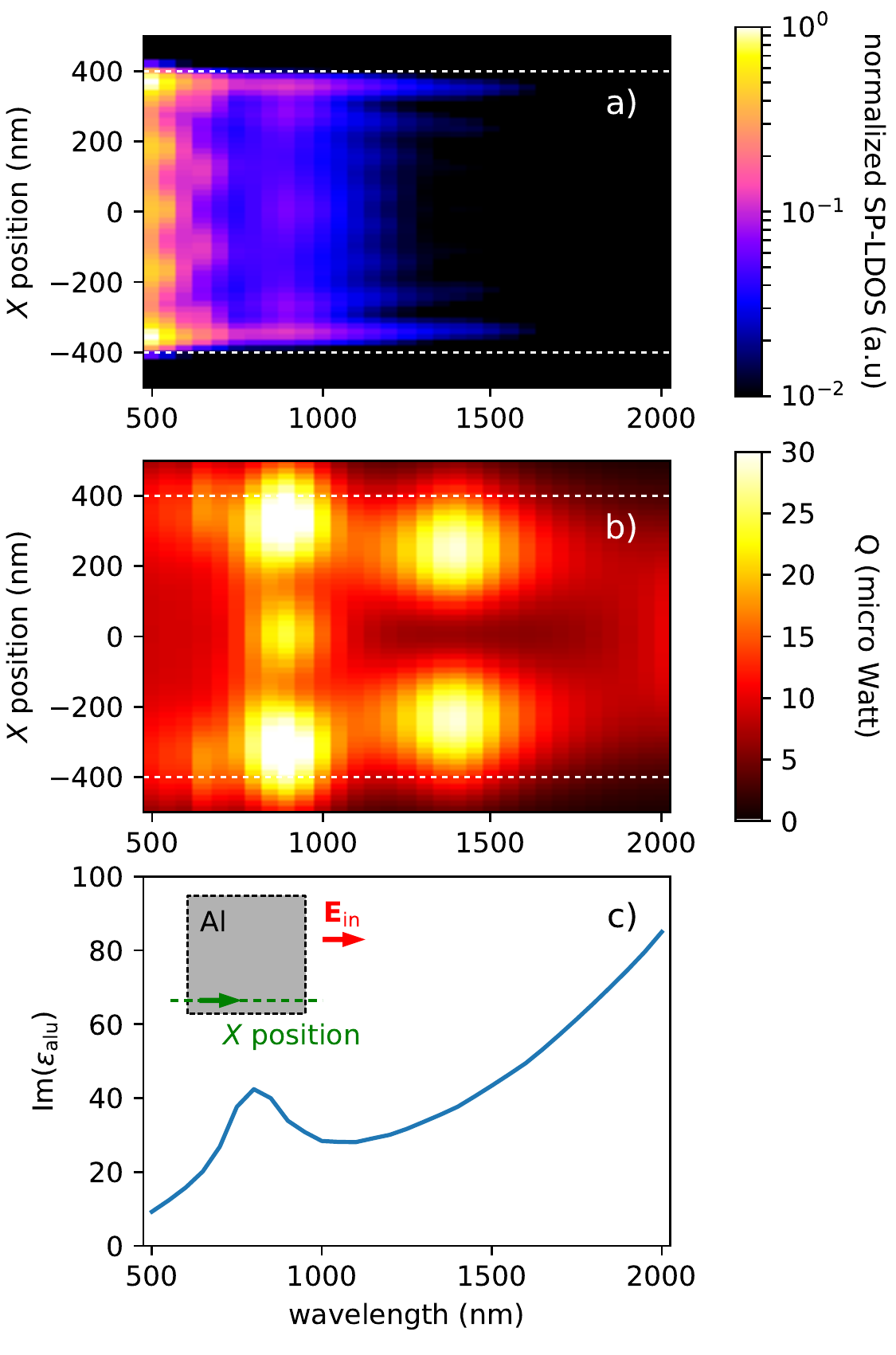}}
\caption{
(a) partial SP-LDOS for polarization along \(X\) and (b) total heat deposited in an \(800\,\)nm Al square on glass substrate, placed in water, as a function of the wavelength and the position of the focal spot. \(X=0\) corresponds to the center of the aluminum pad and its borders are indicated by white dashed lines.
(c) shows the imaginary part of the dielectric function of aluminum.
The angle of the incident linear polarization (\(\mathbf{E}_{\text{in}}\) along \(X\)) and the positions where the data is calculated (green line, \(100\,\)nm above the lower edge of the square) are depicted in the inset in (c).
}
 \label{fig4}
\end{figure}

To assess the influence of these contributions on the heat generation inside large planar aluminum structures, we calculate the partial SP-LDOS along the edge of an aluminum square cavity of \(800\,\)nm side length (\(50\,\)nm height) for light linearly polarized along the square edge.
The Al structure is placed in water, lying on a glass substrate.
We have chosen a square rather than hexagonal cavity because it provides a stronger polarization dependent thermal signature.\cite{SI}
The SP-LDOS and the heat generated in the aluminum square are shown in figure~\ref{fig4}~(a), respectively~(b), as a function of the wavelength and the focal spot position along the square edge.
The considered profile along the Al pad is sketched in the inset in (c). 
In Fig.~\ref{fig4}c the imaginary part of the dielectric function of aluminum is shown.
Following equation~\eqref{eq2} the thermo-plasmonic efficiency is large if the imaginary part of the dielectric function as well as the SP-LDOS (and hence the internal \(\mathbf{E}\)-field) are of high value. 
In the case of the Al square in figure~\ref{fig4}, maximum values for both quantities overlap around \(800\,\)nm for a focused excitation close to the structure borders, where the heat generation is therefore maximal.
At larger wavelengths, the spatial resonances become less pronounced and above \(\approx 1.5\,\)\textmu m no clear resonant behavior is observed in the SP-LDOS.
However, even at these non-resonant wavelengths, there is still a significant heat generation occurring in the aluminum square, which is a result of the large $Im(\epsilon)$, compensating the low electric field intensities inside the object.

\begin{figure}[htb]
\centering{
\includegraphics[width=8cm]{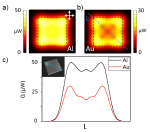}}
 \caption{Reconstructed maps of the total heat deposited in 800 nm (a) Al and (b) Au squares at $\lambda$=750 nm. Each map is the sum of two simulated maps with orthogonal polarizations (white arrows). Simulated (c) crosscuts taken along the blue solid line shown in (b) for the aluminum structure in (a) (red curve) and the gold structure in (b) (black curve).}
 \label{fig5}
\end{figure}

In order to benchmark the efficiency of aluminum for thermal energy conversion, a direct comparison with a similar gold 2D geometry is conducted. Figure \ref{fig5} shows the total heat generation for both an aluminum and a gold square computed for an excitation at $\lambda$=750 nm, corresponding to the wavelength in our nPL experiments and close to the overall maximum heat generation around 800\,nm.
We identify similar features in maps (a) and (b) of Fig.\ref{fig5}. Both structures exhibit an inhomogeneous heat generation landscape with maxima located at each corner of the squares, following the underlying cavity symmetry. These hotspots are connected to each other by a heat background along the edges. A complementary and quantitative information can be extracted from crosscuts along the main diagonal of the square structures (see Fig.\ref{fig5}(b)). While the contrast observed for both metals is driven by the SP spatial distribution, the profiles undoubtedly show that the amplitude of the thermal conversion phenomenon, which is a compromise between the local field $\textbf{E}$ and $Im(\epsilon)$, is increased two-fold in aluminum compared to gold at this specific wavelength. This is understood from the larger dissipation path due to interband transitions in Al in the red part of the visible. Such features are promising for the development and the control of efficient nanosources of heat. 

Once the heat distribution $q(\textbf{R}_{0}, \textbf{r}, \omega)$ deposited in the metal has been computed, the relative increase of temperature $\Delta T$ in the surrounding medium can be easily derived from the expression \cite{Teulle2012}: 

\begin{equation}
\Delta T(\textbf{R}_{0},\omega) = \frac{1}{4 \pi \kappa _{env}} \int_{V} \frac{q(\textbf{R}_{0}, \textbf{r}, \omega)}{\left| \textbf{R}_{0} - \textbf{r} \right|} d\textbf{r}
\label{eq3}
\end{equation}

where $q$ is the power per unit volume dissipated inside the metal, and $\kappa _{env}$ is the thermal conductivity of the immediate surrounding. For $\kappa _{env}$ we took the value associated to water as a residual layer of water is always present at the metal-air interface under ambient experimental conditions. $\left| \textbf{R}_{0} - \textbf{r} \right|$ is the distance between the excitation location $\textbf{R}_{0}$, where the temperature is calculated, and $\textbf{r}$ is an arbitrary location inside the metal. This expression corresponds to a zero-order description of the temperature increase in the structure since it does not take into account the heat diffusion in the metal through the thermal conductivity. A more refined modeling should be considered for the realistic description of extended systems of complex geometries \cite{Baffou2010}. 

\begin{figure}[htb]
\centering{
\includegraphics[width=8cm]{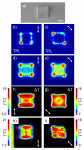}}
\caption{(a) SEM image of a 800 nm aluminum square structure. Scale bar is 300 nm.(b-c) nPL images recorded on this structure for excitation polarization at (b) 90$^{\circ}$ and (c) 45$^{\circ}$. (d-e) Corresponding simulated nPL maps. (f-g) Simulated maps of the temperature variation ($\Delta$T) 150 nm above the metallic square for the two different polarizations. (h-i) are the experimentally reconstructed $\Delta$T maps deduced from the measurement of the resistance of a Au nanowire located at the vicinity of the Al square (not shown) followed by a numerically-assisted calibration step. Polarization orientations are indicated by the white arrows.}
 \label{fig6}
\end{figure}

The SEM image of a 800 nm aluminum square  is shown in Fig.\ref{fig6}(a) with the corresponding experimental (b-c) and simulated nPL maps (d-e) for two different polarizations of the illumination beam. A square belongs to the D$_{4}$ symmetry group;  the optical response of such cavities can therefore be probed by addressing only two preferential symmetry axis corresponding to polarizations separated by 45$^{\circ}$ (here 90$^{\circ}$ and 45$^{\circ}$). The nPL patterns exhibit localized high-intensity regions mainly located at the corners of the structure that are strongly polarization-dependent. Such polarization sensitive response leads to spatial control of the heat generated in the cavity. Remembering that a control over the local temperature in the vicinity of a plasmonic antenna is the cornerstone for trapping and manipulating devices, we compute the pixel-by-pixel rise of temperature 150 nm above the structure by raster scanning the diffraction-limited excitation spot considering a constant incident CW power. 

The results of these numerical experiments are shown in Fig.\ref{fig6}(f) and (g). The $\Delta$T maps display polarization-dependent complex spatial distributions. The temperature increase is only observed when the focused beam is scanned on top of the structure and reaches up to 16 $^{\circ}$C above room temperature for a low cw-equivalent 1 mW excitation at the focal spot (250 nm). For the 45$^{\circ}$ polarization (Fig.\ref{fig6}(g)), there is a temperature increase along the four edges, with maxima located at the four corners and a preferential axis aligned with the incident electric field.  As these $\Delta$T maps are linked to the heat maps and therefore to the experimental nPL ones, the polarization dependence is recovered with the onset of thermal hotspots that fairly follow the spatial distribution of the electric field mapped by nPL microscopy. Recalling that the presence of an interband transition at the laser energy contributes to the amount of heat deposited in the structure, the correlation between the nPL maps and the computed temperature elevation indicates that the plasmonic landscape plays a fundamental role in the photothermal activity enabling thereby a control of the heating efficiency and then the temperature increase by changing the position of the excitation beam with a subwavelength resolution.

In a final step, we quantitatively compare these temperature changes with an experimental $\Delta$T evaluation. The rise of temperature is inferred from a bolometric measure of the electrical transport of a Au nanowire located nearby the aluminum pad. The heat generated by the optical excitation of the high-order SP resonances of the Al cavity diffuses along the substrate interface and affects the temperature-dependent resistance of the nanowire $R_{\rm nw}(T)$ \cite{Bouhelier2016}. A pixel-by-pixel map of the variation of resistance $\Delta R_{\rm nw}$ is thus reconstructed by scanning the Al pad inside the laser focus. The power of the laser before entering the objective is set at 1.1 mW. We then convert the $\Delta R_{\rm nw}$ map into a $\Delta$T map after determining the resistance of the nanowire with a four probe measurement and calibrating the linear rise of the resistance with temperature $R_{\rm nw}(T)$ using a Peltier heating module. This calibration step is completed by a numerical analysis of the geometry using a commercial finite-element software that allows the quantitative estimation of the temperature at any position in the system (the complete procedure is described in reference \citenum{SI}). 
Figures \ref{fig6}(h) and (i) display the experimental $\Delta$T maps deduced from the bolometric measurement for the two polarizations. Although we used a zero-order expression in Eq.\ref{eq3} for the estimation of temperature increase, the same polarization-dependent patterns are observed and the quantitative estimations of $\Delta$T, extrapolated from the numerically-assisted temperature calibration, agree with the simulated ones too. The quantitative agreement confirms the potential of Al to allow for a polarization-controlled heating efficiency leading to a significant temperature increase.  


To conclude, we show that the complex electric field distributions associated to high order plasmonic 
resonances in planar metal cavities may be imaged by nonlinear photoluminescence microscopy in the rather unexplored spectral window around the material's interband transitions. The spatial distribution of the enhanced local electric fields is controlled on-demand by the incident polarization. Our results show that the spatial distribution of the power dissipated in plasmonic nanostructures is governed by the footprint of the supported surface plasmon resonances. We showed, that comparing nPL experiments with GDM simulations can be used as a simple, yet powerful tool to predict the thermoplasmonic properties of complex multi-modal 2D aluminum structures. Numerical simulations show that aluminium can yield a thermoplasmonic response larger than gold in the infrared spectral range due to its interband transitions. This analysis directly leads to the quantitative estimation of the temperature increase in the direct vicinity of an Al structure which is in excellent agreement with measured $\Delta$T. For a moderate excitation power, the relative temperature increase for a specific excitation location can reach 15-20 $^{\circ}$C at a vertical distance of 150 nm in the surrounding. The spectral overlapping of high order SP resonances with the interband transitions in such aluminum objects results in a controlled heat generation and corresponding on-demand temperature increase in the near field. Tailoring of the spatial heat generation can be achieved by adjusting the cavity geometry and the incident polarization. Our results pave the way to new experiments and applications in thermoplasmonics where stronger photothermal effects are required, as well as to new strategies for thermo-optical nanomanipulation, hyperthermia-induced apoptosis in nanomedecine or thermally activated nanochemistry. 

\indent {\it Acknowledgments -} 

We are grateful to Christian Girard for discussions and for critical reading. This work was supported by the CPER ``Gaston Dupouy'' 2007-2013 and the massively parallel computing center CALMIP in Toulouse. This work has been partially funded  by the European Research Council under the European Community's Seventh Framework Program FP7/ 2007-2013 Grant Agreement No. 306772. Samples were fabricated using the Nanomat nanofabrication facility (www.nanomat.eu) and the ARCEN-Carnot Platforme. DK acknowledges support from Region Champagne-Ardenne.


\clearpage
\widetext

\section*{Supporting informations: Local field enhancement and thermoplasmonics in multimodal Aluminum structures}
%
%
\setcounter{equation}{0}
\setcounter{figure}{0}
\setcounter{table}{0}
\setcounter{page}{1}
\setcounter{section}{0}
\makeatletter
\renewcommand{\theequation}{S.\arabic{equation}}
\renewcommand{\thefigure}{S.\arabic{figure}}
\renewcommand{\bibnumfmt}[1]{[S#1]}
\renewcommand{\citenumfont}[1]{S#1}

\begin{figure}[ht!]
\centerline{\includegraphics[width=10cm]{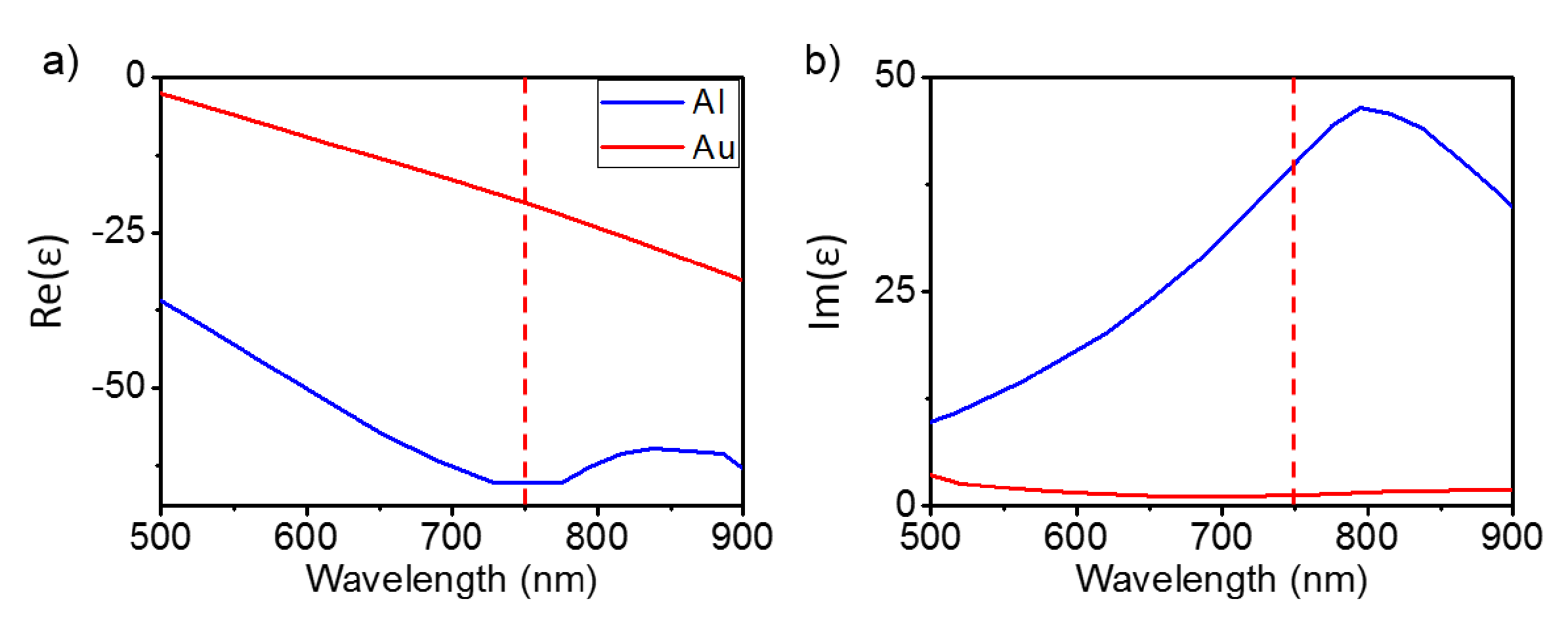}}
\caption{Real (a) and imaginary (b) part of the dielectric function of gold (red line) and aluminum (blue line).}
\label{FigS1}
\end{figure}

\begin{figure}[ht!]
\centerline{\includegraphics[width=\textwidth]{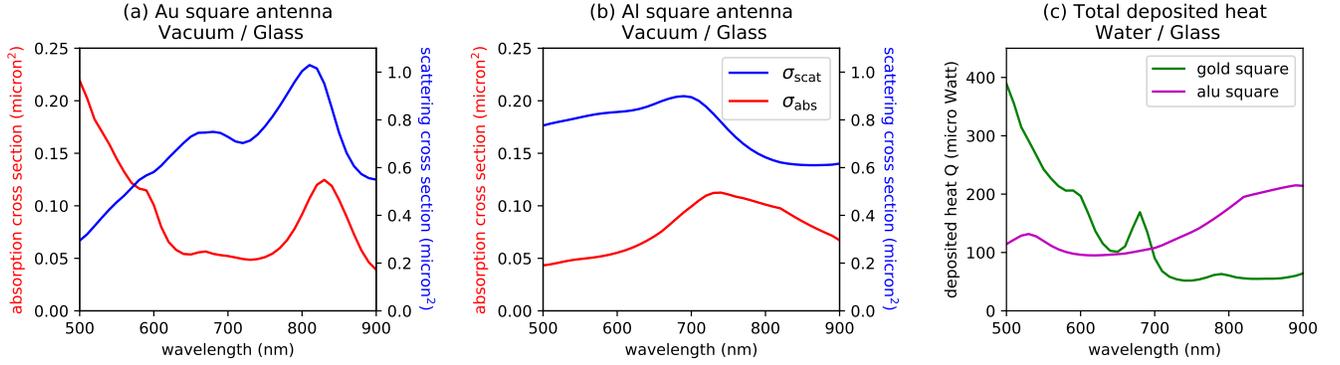}}
\caption{Scattering (blue) and absorption (red) spectra for a square pad antenna (side length~\(800\,\)nm, height~\(50\,\)nm) placed in vacuum, made from (a) gold and (b) aluminum.
(c) total deposited heat in the square antenna surrounded by water (gold: green, aluminum: pink). 
The square antennas are placed on a glass substrate.}
\label{FigS1b}
\end{figure}

\begin{figure}[ht!]
\centerline{\includegraphics[width=8cm]{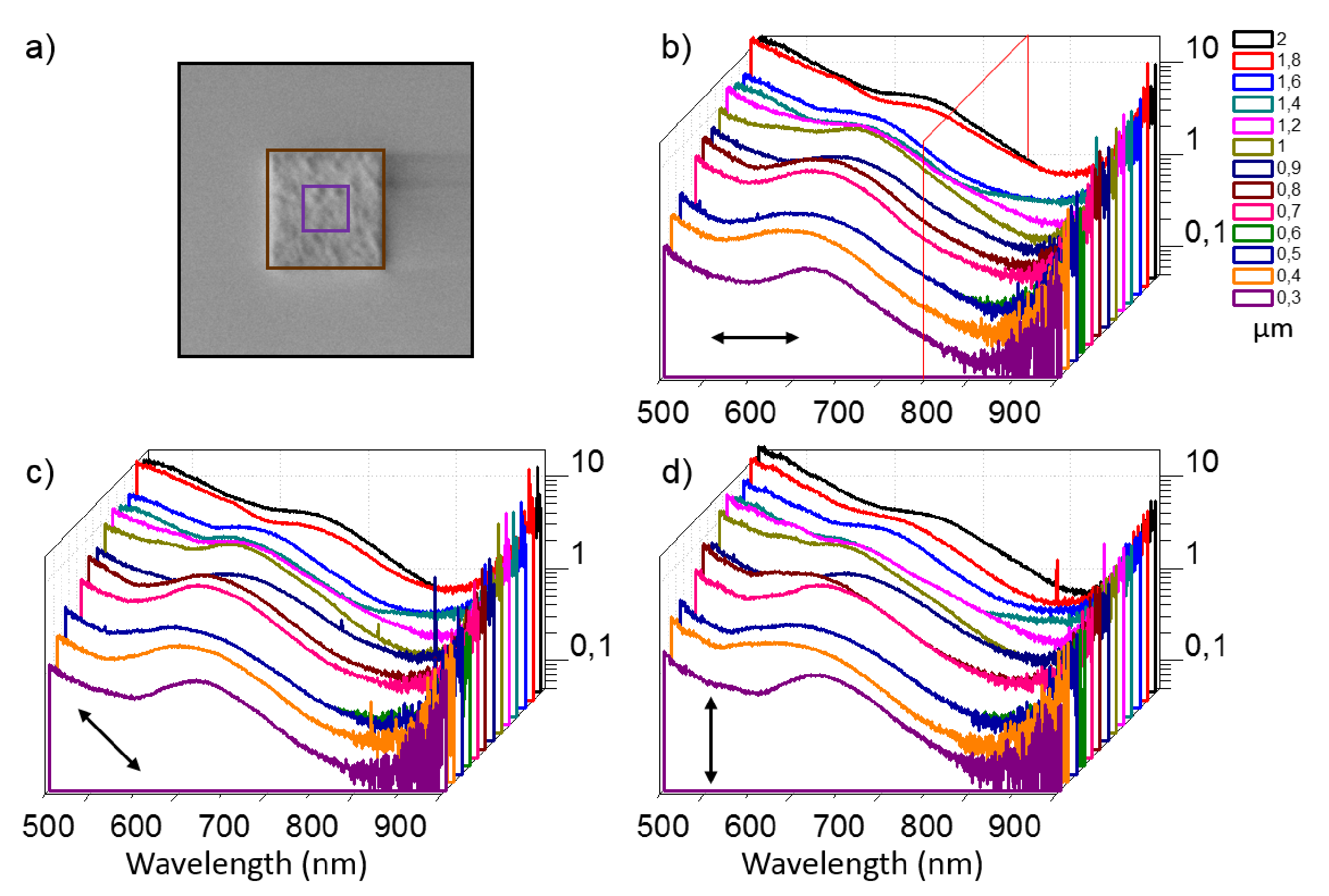}}
\caption{(a) SEM image of a 800 nm aluminum square. The violet and the black lines correspond respectively to a 300 nm and a 2 $\mu$m structure. Dark-field spectra acquired on aluminum square structures with side lengths ranging from 300 nm to 2 $\mu$m for linearly polarized excitation at (b) 0$^{\circ}$, (c) 45$^{\circ}$ and (d) 90$^{\circ}$. The experimental nPL excitation wavelength at 750 nm is indicated in (b) by a red line.}
\label{FigS2}
\end{figure}

\section{Scattering and absorption properties of Al nano- and micro-structures (Figures S\ref{FigS1}, S\ref{FigS1b} and S\ref{FigS2})}

One of the two channels for temperature increase identified in the main text is the intrinsic dissipative nature of the metal, depicted by the imaginary part of the dielectric function Im$(\epsilon)$. Figure S\ref{FigS1} shows a comparison of the real and imaginary part of the dielectric function for gold and aluminum. As a result of a higher carrier density compared to gold and silver, aluminum features a high plasma frequency. For this reason, Al opens a new field of research for UV-plasmonics. Furthermore it displays strong interband transitions in the visible with a maximum at 800 nm, resulting in a maximum Im$(\epsilon)$, as shown in Fig.S\ref{FigS1}(b) \cite{Gerard2015}. This behavior is of particular interest for thermoplasmonics, since the photothermal energy conversion is driven by Ohmic losses in the metal.

Figure~\ref{FigS1b} shows simulated scattering and absorption spectra for a square antenna of \(800\,\)nm side length and \(50\,\)nm height, placed on a glass substrate (\(n_{\text{glass}}=1.45\)) in vacuum (\(n_{\text{vac}}=1.0\)). 
The pad is illuminated by a plane wave, polarized along the square side (along \(X\)).
In (a) the square antenna is made from gold, while in (b) aluminum is used as material.
Within the limits of the simulation's approximations (plane waves, total scattering) the scattering peak around~\(700\,\)nm for the Al square matches the experimental scattering spectra (see below).
As for the absorption, we observe for gold a strong increase of \(\sigma_{\text{abs.}}\) occurring below \(\approx 600\,\)nm, where simultaneously scattering decreases.
For the aluminum pad on the other hand, the absorption increases rather towards the ``red'' part of the spectral range. 

The deposited heat is shown in Fig.~\ref{FigS1b}c for the square of either gold (green line) or aluminum (pink line), placed on a glass substrate in water (\(n_{\text{water}}=1.33\), similarly to simulations in the main text), again under plane wave illumination polarized along \(X\).
The deposited heat follows the trends of the absorption spectra, which is an expected result.
Differences are mainly a consequence of the water environment (vacuum for the scattering and absorption spectra).
Comparing the heat deposition in a gold and an aluminum square, we observe that the gold antenna generates a strong amount of heat in the visible spectral range, where it clearly outperforms the Al antenna in terms of energy absorption.
However, above \(\approx 700\,\)nm, the aluminum antenna generates more heat than its gold counterpart, outperforming it in the near infrared by up to a factor of~\(\approx 3\).

Experimental dark-field scattering spectra from aluminum square structures, recorded for several incident polarizations, are gathered in Fig.S\ref{FigS2}. The spectra exhibit all the same trend, regardless their side lengths which range from 300 nm to 2 $\mu$m in our sample. Due to the maximum dissipation (\(\mathsf{Im}(\epsilon)\)), there is a minimum of the scattered signal around 800 nm, independently of the structure size or the incident polarization.

\newpage

\begin{figure}[ht!]
\centerline{\includegraphics[width=8cm]{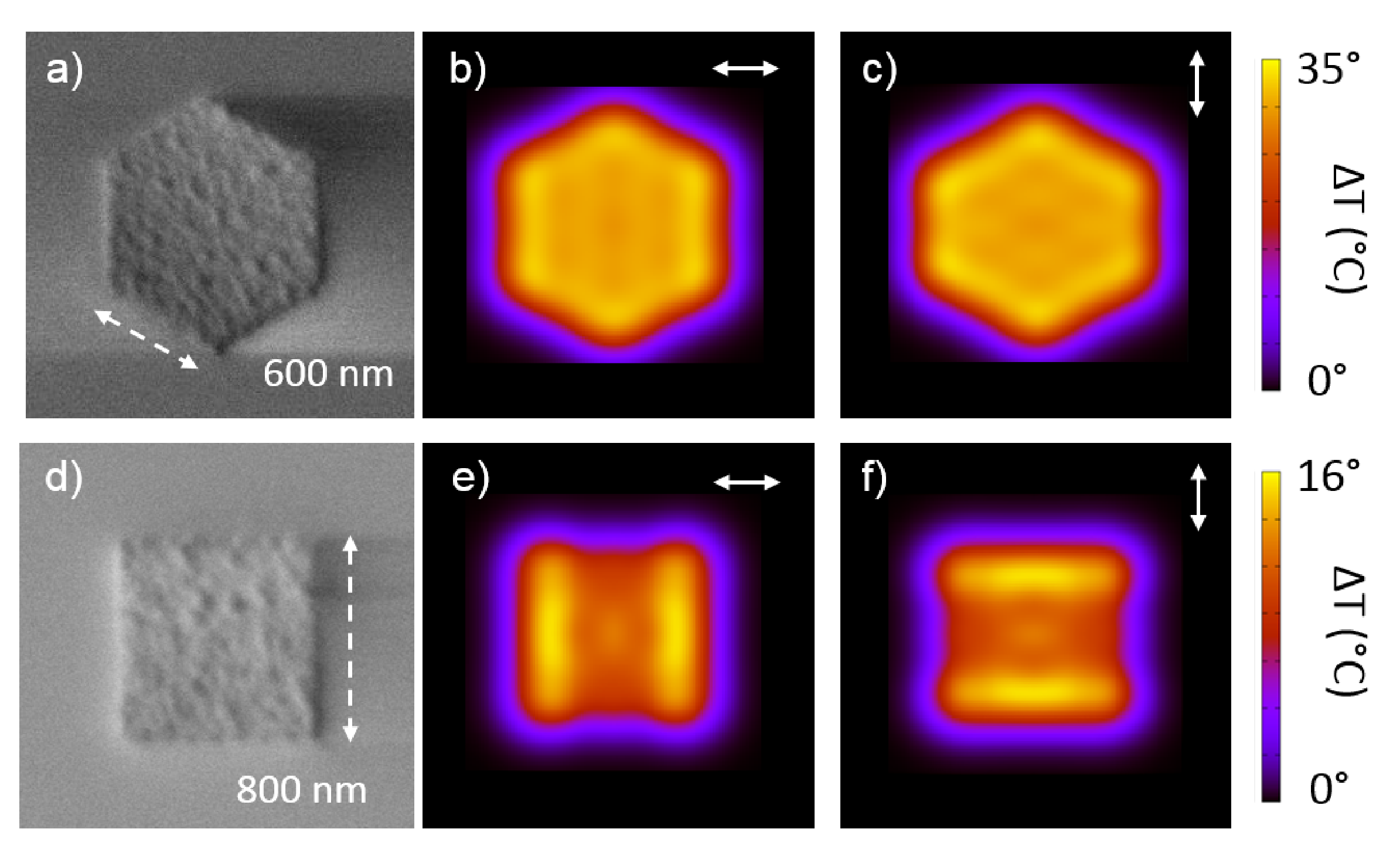}}
\caption{(a) SEM image of a 600 nm aluminum hexagonal structure. (b-c) Simulated maps of the temperature variation ($\Delta$T) above the metallic hexagon for two different polarizations: (b) 0$^{\circ}$ and (c) 90$^{\circ}$. (d) SEM image of a 800 nm aluminum square structure. (e-f) Simulated maps of the temperature variation ($\Delta$T) above the metallic square for two different polarizations: (e) 0$^{\circ}$ and (f) 90$^{\circ}$. Polarization orientations are indicated by the white arrows. Raster-scan simulations using a focused incidence.}
\label{FigS3}
\end{figure}

\section{Optimum geometry for polarization-dependent temperature increase (Figure S\ref{FigS3})}

Both hexagonal and square geometries were investigated in this work. The nPL experiments demonstrated that several hotspots can be generated and controlled by changing the incident polarization. The nPL signal provides a direct information on the local field enhancement. Conclusions on the local absorption, and therefore on the heat generation can be drawn from this information, because both mechanisms are linked to the local electric field. The temperature maps presented in Fig.S\ref{FigS3} are simulated for a square and a hexagon structure with respective side lengths of 800 and 600 nm. These results are representative for what was systematically observed in numerical experiments. Under focused illumination, the hexagonal geometry shows the greatest temperature elevation (up to 35$^{\circ}$C above room temperature with an excitation density of 0.1 MW.cm$^{-2}$). However, squares display a higher contrast in the $\Delta$T landscape as well as a stronger polarization dependence. The resulting temperature gradients can be controlled by rotating the incident polarization, a functionality of particular interest for thermo-optical trapping and nanomanipulation in fluidic environments. These features motivated the choice of the square geometry for the discussion on temperature increase in the main text.

\newpage

\begin{figure}[ht!]
\centerline{\includegraphics[width=8cm]{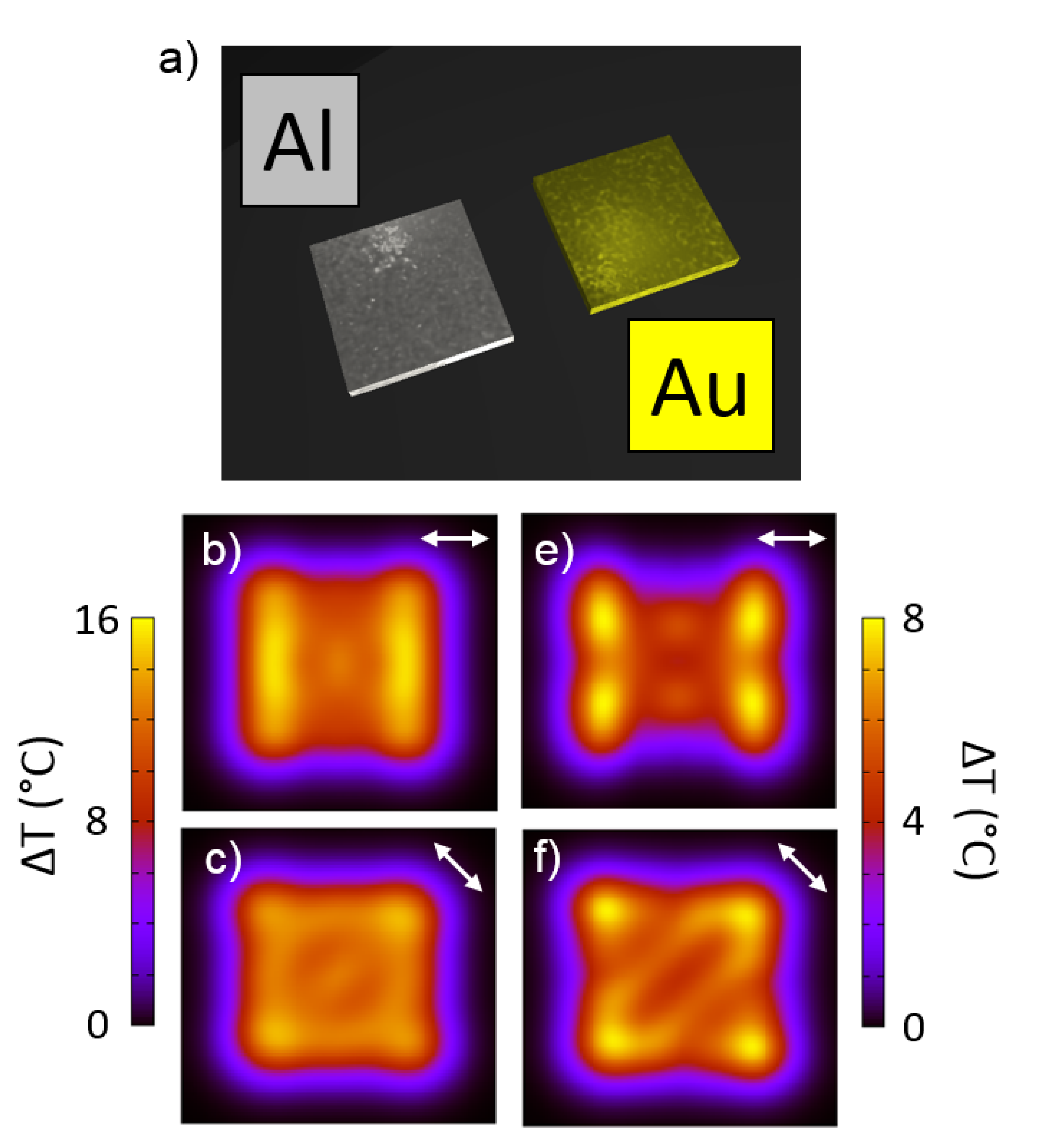}}
\caption{(a) Artistic view of an aluminum and a gold square structure. Simulated map of the temperature increase 150 nm above a 800 nm Al square at $\lambda$=750 nm for a focused excitation at (b) 0$^{\circ}$ and (c) 45$^{\circ}$ (P=1mW). Simulated map of the temperature increase above a 800 nm Au square at $\lambda$=750 nm for a focused excitation at (e) 0$^{\circ}$ and (f) 45$^{\circ}$.}
\label{FigS4}
\end{figure}

\section{Comparison of temperature elevation in Au and Al structures (Figure S\ref{FigS4})}

The $\Delta$T maps simulated at $\lambda$=750 nm are shown in Fig.S\ref{FigS4} above an aluminum and a gold square (cw excitation density of 0.1 MW.cm$^{-2}$). While the different dielectric constants imply plasmonic modes of different orders at a given energy, the $\Delta$T patterns are quite similar with the same dependence to the polarization.

Nevertheless, we want to emphasize two important differences. First, in the case of the gold structure more distinct features are obtained in the $\Delta$T maps compared to Al. Second, the maximum increase of temperature is twice as high in the vicinity of the Al structure than for the Au one (16$^{\circ}$C compared to 8$^{\circ}$C). These two observations rely on the subtle compromise between spatial mode distribution in the cavity and imaginary part of the dielectric function, as discussed in the main paper. The strong temperature increase expected for such structures makes aluminum a promising alternative for thermo-plasmonics experiments.

\newpage

\begin{figure}[ht!]
\centerline{\includegraphics[width=8cm]{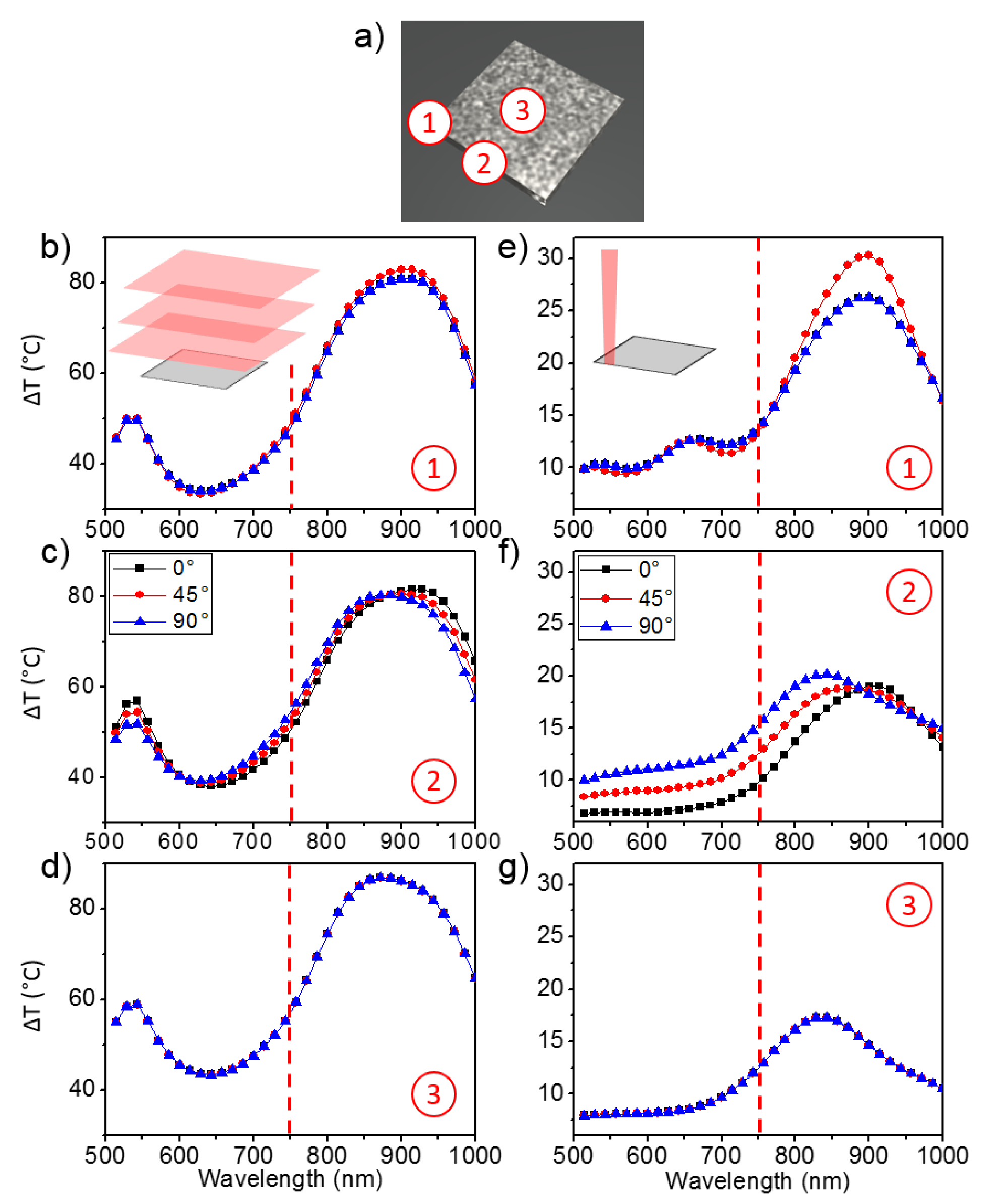}}
\caption{(a) Artistic view of Al square structure, with indications of the positions of three probing positions. (b-d) Polarization-dependent temperature increase as a function of the wavelength simulated above the aluminum structure illuminated by a plane wave. $\Delta$T is computed at the positions labeled 1 (b), 2 (c) and 3 (d) in (a). (e-g) Same as (b-d) for a focused illumination parked in 1 (e), 2 (f) or 3 (g).}
\label{FigS5}
\end{figure}

\section{Temperature evolution above an Al structure as a function of the wavelength and the nature of the illumination (Figure S\ref{FigS5})}

In figure S\ref{FigS5} the simulated temperature evolution 150 nm above a 800 nm aluminum square is shown as a function of the wavelength, the polarization and the illumination mode (focused beam or plane wave). In the case of a plane wave illumination, the temperature increase as a function of the wavelength is sensitive neither to the probing position, nor to the polarization of the incident light. 
In contrast, the focused excitation gives rise to different responses depending on the position of the illumination spot. Once the spot is parked, a rotation of the polarization also permits a local modulation of the temperature increase. One can note that this dependence to the orientation of the electric field is not constant over the spectral window explored here. 
Independently of the illumination mode, both configurations show a similar trend. As expected from the imaginary part of the dielectric function, the heat generation and the subsequent temperature increase are stronger when the illumination wavelength approaches 800 nm. Interestingly, the largest temperature variation is obtained in the 850-950 nm spectral window and not at 800 nm.

Finally we note that the temperature elevation is much stronger in the case of a plane wave illumination (up to 80$^{\circ}$C above room temperature) for a fixed incident excitation density of 0.1 MW.cm$^{-2}$ for both illumination modes. 

\newpage

\begin{figure}[ht!]
\centerline{\includegraphics[width=15cm]{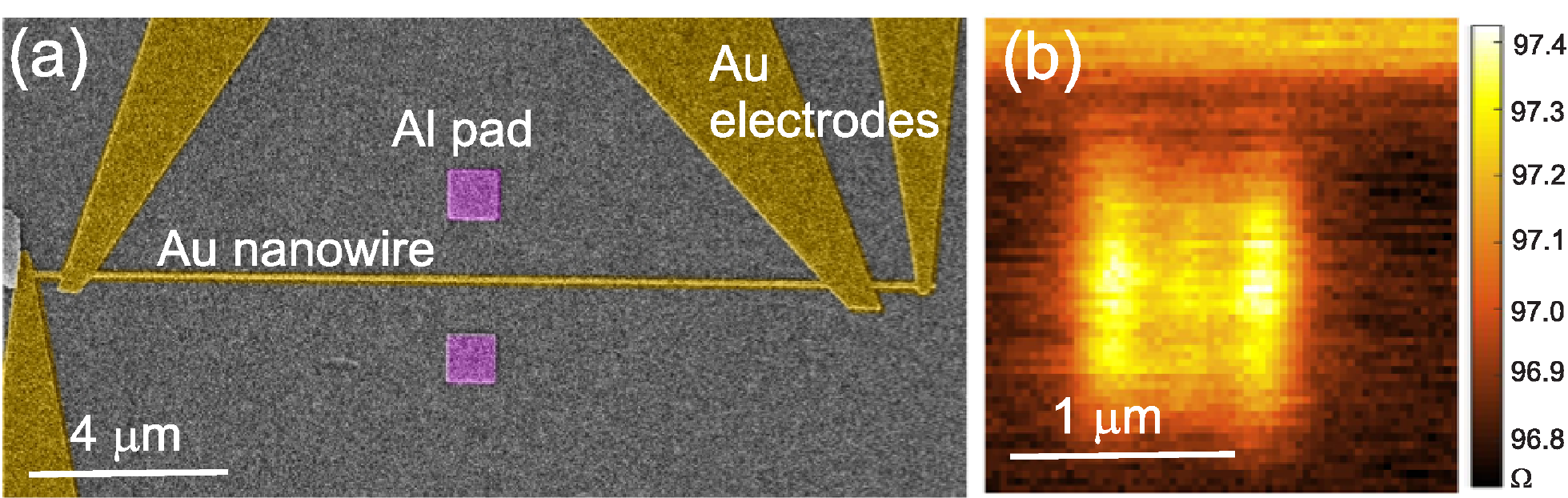}}
\caption{(a) Scanning electron micrograph showing two Al pads placed adjacent to a Au nanowire. The set of Au electrodes for determining the resistance of the nanowire in a four-probe configurations are also visible. (b) Reconstructed map of the variation of the nanowire resistance when the Al pad is scanned in the focus of the laser beam. The incident polarization is horizontal.}
\label{FigS6}
\end{figure}

\begin{figure}[ht!]
\centerline{\includegraphics[width=15cm]{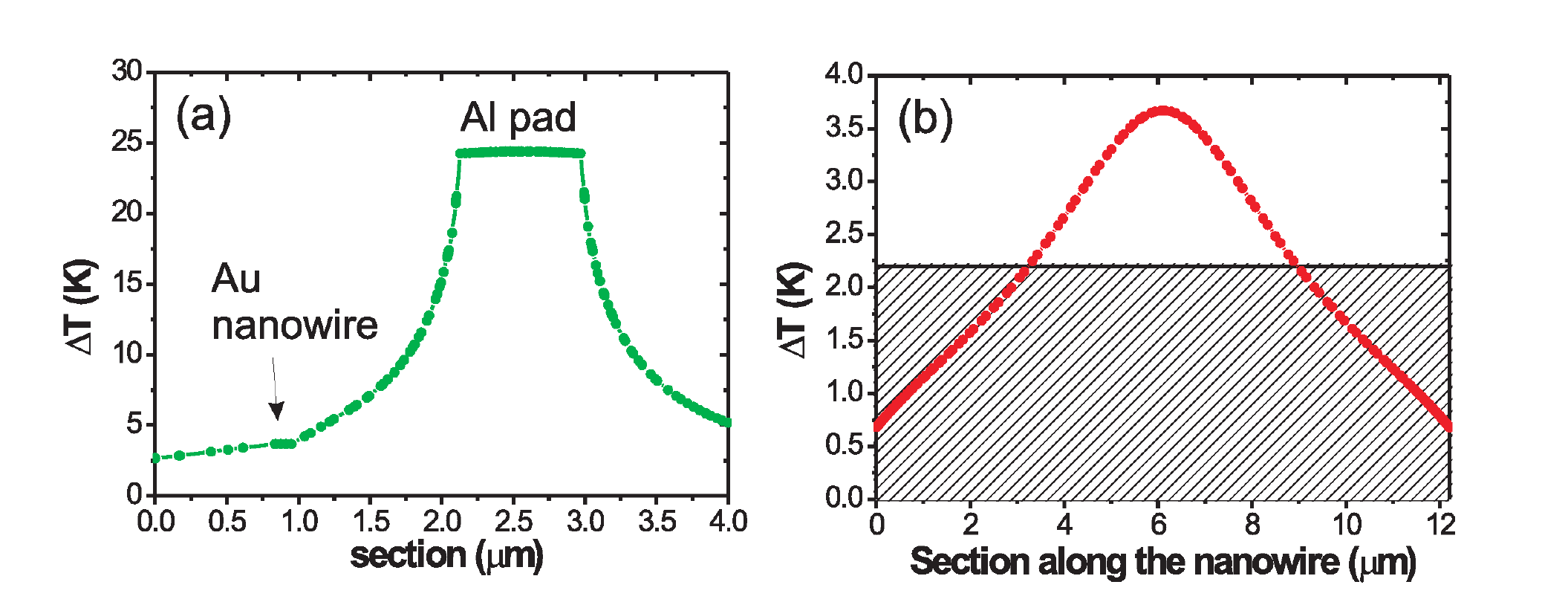}}
\caption{(a) Diffusion of the temperature in the calculation window considering a homogeneously heated Aluminum pad. The presence of the nanowire (arrow) introduces a step in the diffusion cross-cut. (b) Profile of the temperature taken along the nanowire. The constant line represents the constant temperature of the nanowire considering a global heating generating a 0.8$\Omega$ change of resistance. The points are showing the calculated temperature profile considering a local heating depicted by (a). The areas behind the two curves are equal when the temperature of the Al pad in (a) is set at 24.4$^{\circ}$K above room temperature.}
\label{FigS7}
\end{figure}

\section{Bolometric determination of the temperature (Figures S\ref{FigS6} and S\ref{FigS7})}

In this section we briefly describe the procedure applied to quantitatively estimate the temperature rise of the Al pads. The approach relies on measuring the electronic transport of a Au nanowire placed at the immediate vicinity of the Al structure. When the Al pad is illuminated, the heat generated upon absorption diffuses through the substrate and raise the temperature of the Au nanowire. We monitor the temperature-dependent resistance $R_{\rm nw} (T,x,y)$ of the nanowire while the Al pad is scanned pixel by pixel in the focus of the 785 nm CW laser beam.
The aluminum pad, the gold nanowire and the electrical leads connecting the extremities of the nanowire are realized by three steps of lithography. The nanowire and the set of proximity electrodes are produced by an electron-beam lithography routine including development, metal evaporation, and liftoff of the electron-sensitive resist. A second electron-beam lithography routine, aligned with the first one, defines the Al pads with respect to the nanowire. The thickness of the Au layer and of the Al pads are fixed at $\sim$ 50 nm. A final optical lithography is performed to produce the microscopic electrodes connecting the nanowire to an outside power supply and signal recovery equipment. Figure~S\ref{FigS6}(a) shows a scanning electron micrograph of a typical sample. Two 830~nm squared Al pads are placed at 930~nm from the edge of the nanowire. The latter is 12.2 $\mu$m long and 110~nm wide. We use a four probe configuration to determine the resistance of the nanowire following the procedure described in \cite{Bouhelier2016}. The resistance of the nanowire measured at room temperature is here 96.4~$\Omega$.

Figure~S\ref{FigS6}(b) shows the variation of the resistance of the nanowire   $R_{\rm nw}$ measured while the bottom Al pad is scanned through the focus of 1.1 mW laser beam focused by a high numerical objective (N.A.=1.49). The polarization of the incident beam is horizontal in this example. When the laser overlaps the metal parts of the sample, the heat produced by absorption induces an elevation of the resistance that is recorded pixel by pixel. The largest variation measured here is $\sim$0.8$\Omega$. 

We then proceed to calibrate the change of the resistance with temperature by placing the entire sample on a voltage-controlled Peltier module. We measure the variation over a temperature range spanning from room temperature up to 450~K and find a linear coefficient of 0.32~$\Omega$K$^{-1}$. As discussed in \cite{Bouhelier2016}, a calibration solely relying on a global heating of the entire substrate underestimates the local elevation of the temperature generated at the laser focus. To retrieve a quantitative estimation of the temperature at the position of the Al pad from the resistance map, we perform a numerical analysis of the geometry using a commercial finite-element software (Comsol multiphysics) taking into account the experimental conditions. The three-dimensional calculation window is composed of a glass substrate and an air superstrate. We model the Al pad as a homogeneous heated area and simulate the heat diffusion in the system. Figure~S\ref{FigS7}(a) shows a cross-cut of the temperature gradient taken perpendicular to the nanowire and passing through the Al structure. The temperature of the Al pad is constant and decays as heat diffuses away. To determine the value of the temperature of the heated pad, we proceed as follow: the highest variation of resistance is measured at 0.8$\Omega$ (Fig.~S\ref{FigS6}(b)). Using the calibrated temperature coefficient, this would correspond to a homogeneous temperature of the nanowire of 2.2 K.  This is represented by the constant line in Fig.~S\ref{FigS7}(b) showing the temperature profile along the nanowire. The data points are displaying the calculated temperature profile along the nanowire considering a local heating, \textit{ i.e.} the situation depicted in Fig.~S\ref{FigS7}(a). We then equalize the areas behind the two curves to determine the temperature of the Al pad. In other words, we set equal the resistance variation measured by  a global heating of the nanowire using the Peltier module and the resistance change measured by a local heating of the Al pad. For the present case, the areas are equal if the temperature of the Al pad is set at 24.4~K above room temperature. The calculation shows that at 150~nm above the structure, the temperature decreases to 24.3~K above room temperature.

\newpage

\begin{figure}[ht!]
\centerline{\includegraphics[width=8cm]{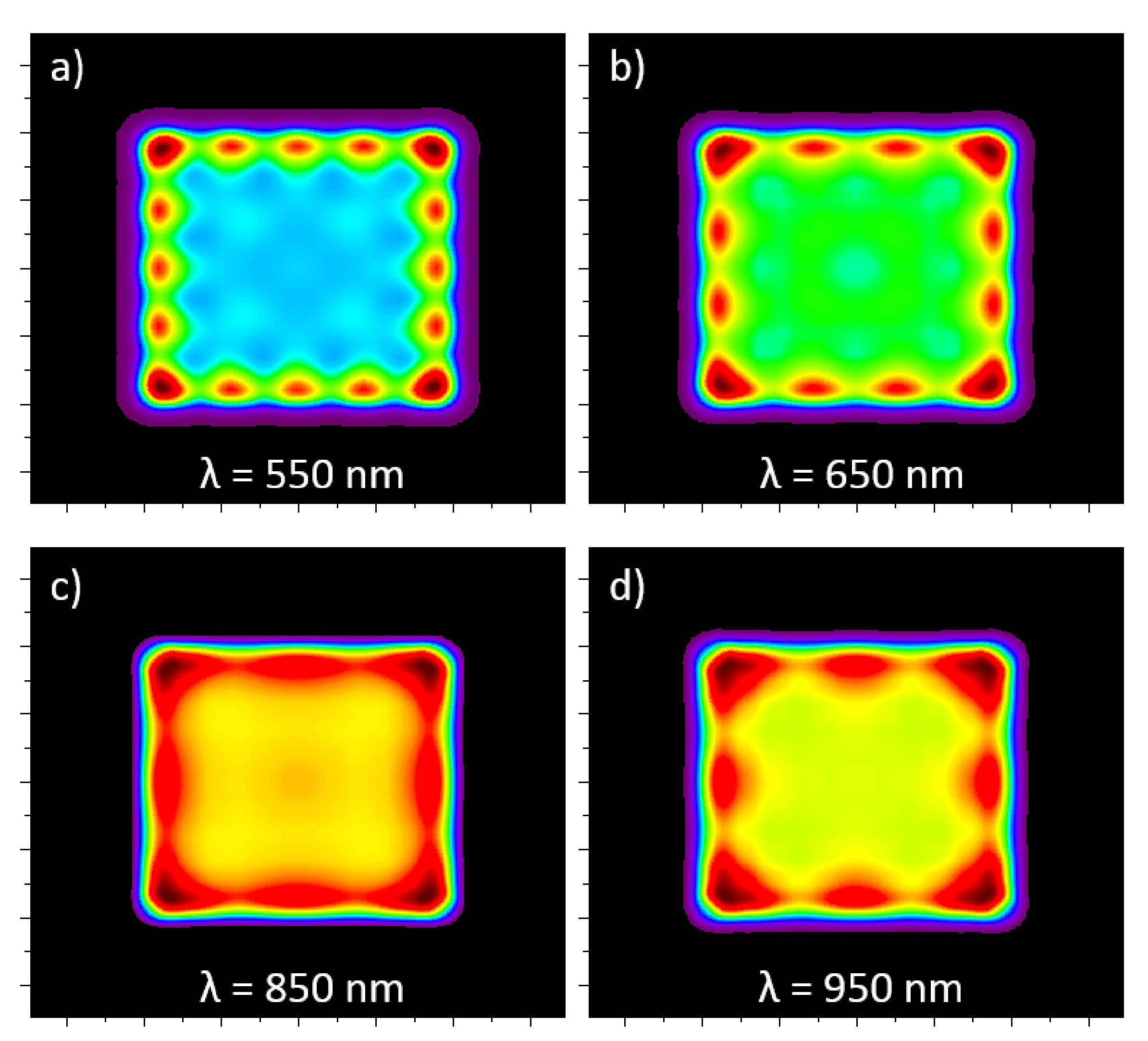}}
\caption{SP density of states (SP-LDOS) maps computed for the 800 nm square in Aluminum at (a) 550 nm, (b) 650 nm, (c) 850 nm and (d) 950 nm.}
\label{FigS8}
\end{figure}

\section{Comparison of the SP-LDOS spatial distribution as a function of the wavelength (Figure S\ref{FigS8})}

In figure S\ref{FigS8} the simulated SP-LDOS spatial distribution in a 800 nm Aluminum square is shown as a function of the wavelength, complementary to figure 1 (b) of the main manuscript. One can clearly see the evolution of the number of lobes along the edges from 1 to 3 when the energy is increased. These specific spatial distributions are the signature of the high-order plasmonic resonances sustained by the cavity.

\end{document}